\documentclass[prl,nofootinbib]{revtex4}
\usepackage{graphicx}
\usepackage{latexsym}

\newcommand{\be}{\begin{equation}}
\newcommand{\ee}{\end{equation}}
\newcommand{\beqa}{\begin{eqnarray}}
\newcommand{\eeqa}{\end{eqnarray}}

\begin{document}
\title{Probing for Variation of Neutrino Mass with Current Observations}

\author{Gong-Bo Zhao}
\author{Jun-Qing Xia}
\author{Xinmin Zhang}

\affiliation{Institute of High Energy Physics, Chinese Academy of
Science, P.O. Box 918-4, Beijing 100049, P. R. China}

\begin{abstract}
With the latest astronomical data including Cosmic Microwave
Background (WMAP three year, CBI, ACBAR, VSA), Type Ia Supernova
(``gold sample"), Galaxy Clustering (SDSS 3-D matter power,
Lyman-$\alpha$ forest and Baryon Acoustic Oscillating (BAO)), we
make a global fitting to constrain the mass varying neutrinos. We
find that the parameter $\delta$, denoting time evolving of
neutrino mass, is weakly constrained and the neutrino mass limit
today can be relaxed at least by a factor of two. Adding data of
$0\nu2\beta$ decay of Heidelberg-Moscow experiment to our
analysis, we find that $\delta$ can be constrained tightly and
mass varying neutrinos are favored at about $99.7\%$ confidence
level.
\end{abstract}

\maketitle

The mass varying neutrino which has been studied actively in the
literature recently brings forth new theoretical challenges and
abundant phenomenology. The predictions of the variation of the
neutrino masses can be tested in the experiments of neutrino
oscillation\cite{nuosc}, short gamma ray bursts\cite{Li:2004tq} and
extremely high energy cosmic neutrino\cite{ring}. Cosmologically the
mass varying neutrino plays an interesting role in determining the
evolution of the universe\cite{Bi:2004ns} and has interesting
implications in leptogenesis\cite{Bi:2003yr},
Supernova\cite{Li:2005zd}, the Cosmic Microwave Background(CMB) and
Large Scale Structure(LSS)\cite{Brookfield:2005td}.

Neutrino mass variation can be induced by the interaction between
neutrino and scalar field\cite{Gu:2003er,Fardon:2003eh,
peccei,Zhang:2005yw} in models of neutrino dark energy\footnote{
In Ref.\cite{Afshordi:2005ym} Afshordi \emph{et al.} concluded the
neutrino dark energy model presented by Fardon \emph{et
al.}\cite{Fardon:2003eh} is unstable due to the negative sound
speed squared. However, their argument is based on the \emph{ad
hoc} model in \cite{Fardon:2003eh} thus lack the generality.
Generically, for the mass varying neutrino model originally
proposed in \cite{Gu:2003er}, there exists large parameter space
where the system is stable. For a very recent survey, see
Ref.\cite{Bjaelde:2007ki}.}. In the framework of $\Lambda$CDM the
variation of neutrino mass could result from the coupling between
neutrino and Ricci scalar in the form of $f(R)\bar \nu \nu$ and
possibly some other unknown mechanism. In this paper we work in a
model-independent way to constrain the time evolving of neutrino
mass by parameterizing neutrino mass in the form of
Eq.(\ref{para}). For simplicity and to extract the effect of mass
varying neutrinos we use the simplest dark energy model, the
cosmological constant for the analysis in this paper.

We take a global fit to constrain the parameter $\delta$ denoting
the mass variation of neutrinos and study its effect on the
determination of other cosmological parameters. Our result shows
that 1)with the mass variation, the cosmological limit on the
neutrino mass is relaxed by roughly a factor of two; 2)if taking
the limits on the neutrino mass from Heidelberg-Moscow
experiment(HM) as a prior\cite{Kl04,Kl06,fogli2}, we find that the
variation of neutrino mass is tightly constrained and the mass
varying neutrinos are favored at about 3 $\sigma$.

 We assume that three species of neutrinos whose masses are
 degenerate
and parameterize the neutrino mass in form of \footnote{
     The
     physical motivation for mass varying neutrino mainly stems from the coupling of
     neutrino and
     the dark energy thus we choose this parametrization which of similar form of
     widely-used parametrization for dark energy, namely, $w(a)=w_0+w_1(1-a)$.
     Here  $m_{\nu 0}$ and $\delta$ denote the neutrino mass presently and its derivative
     with respect to scale factor, say, $dm(a)/da$. In this sense, we work in a model-independent
     fashion. We stress that our parametric method using Eq.(\ref{para}) is just one possible
     way to study the variation of neutrino mass. With the
     accumulation of high quality astronomical data, we can
     investigate the neutrino mass variation in a more
     model-independent way, such as using the technique of
     Principle Component Analysis(PCA).
      }
\begin{equation}\label{para}
     m_{\nu}(a)= m_{\nu 0}[1+\delta(1-a)]
\end{equation}
where $ m_{\nu}(a)$ is the sum of neutrino masses which is a
function of scale factor $a$, $ m_{\nu 0}$ is the sum of neutrino
mass at present, $\delta$ is the dimensionless factor denoting the
time-varying effect of neutrino masses. For the background,  \be
\rho_\nu = \frac{1}{a^4} \int q^2 dq d\Omega \epsilon f_0(q), \ee
with $\epsilon^2 = q^2 + m^2_\nu(a) a^2$, $q^i = a p^i$ is the
comoving momentum. The pressure is \be p_\nu = \frac{1}{3 a^4}
\int q^2 dq d\Omega f_0(q) \frac{q^2}{\epsilon}. \ee Thus, \be
\dot\rho_\nu + 3H(\rho_\nu + p_\nu) = \frac{\partial \ln
m_\nu(a)}{\partial a}\dot a(\rho_\nu - 3p_\nu). \ee where the
overdot represents the derivative with respect to the conformal
time. For the evolution of perturbation, working in the
synchronous gauge and following the treatment of
Ref.\cite{Ma:1995ey}, we find that the Boltzmann equations Eq.(40)
in \cite{Ma:1995ey} don't relate directly to the mass varying
terms $\partial \ln m_\nu/\partial a$, and therefore the hierarchy
equations for the massive neutrinos Eq.(56) in \cite{Ma:1995ey}
remain unchanged. In our numerical calculation we integrate these
hierarchy equations rather than use the approximate strategy
suggested in \cite{Lewis:2002nc} to assure a high precision. Using
a modified code of \texttt{CosmoMC}\cite{Lewis:2002ah},
we make a global fit to constrain
the mass varying neutrinos. Assuming a flat universe, we set the
most general cosmological parameter space as:
\begin{equation}\label{param}
    \textbf{P}\equiv(\omega_{b}, \omega_{c}, \Theta_S, \tau,  f_\nu, \delta, n_{s}, \log[10^{10}
    A_{s}])~,
\end{equation}
where $\omega_{b}=\Omega_{b}h^{2}$ and
$\omega_{c}=\Omega_{c}h^{2}$ are the physical baryon and cold dark
matter densities relative to the critical density, $\Theta_S$
characterizes the angular scale of sound horizon, $\tau$ is the
optical depth to the last scattering surface and $A_{s}$ is
defined as the amplitude of initial power spectrum, $f_\nu$ is the
dark matter neutrino fraction at present, namely,
\begin{equation}\label{fnu}
    f_\nu =\frac{\rho_{\nu}}{\rho_{DM}}=\frac{
    m_{\nu 0}}{93.105~eV~\Omega_{c}h^{2}}~~,
\end{equation}
where  $\rho_{\nu}$ and $\rho_{DM}$ denote the energy density of
neutrino and dark matter at present respectively. We sample in
above an eight dimensional parameter space and fit the theoretical
output to the observation using Markov Chain Monte Carlo
algorithm\cite{MCMC97,MacKayBook,Neil93}. We
take the weak priors as: 
$\tau<0.8$, $0.5<n_{s}<1.5$, $0<f_{\nu}<0.5$, a cosmic age tophat
prior as 10 Gyr$<t_{0}<$20 Gyr. To keep the positivity of neutrino
mass we take $-1<\delta<10$. Furthermore, we make use of the
Hubble space telescope (HST) measurement of the Hubble parameter
$H_0 = 100h ~\text{km s}^{-1} \text{Mpc}^{-1}$ \cite{freedman} by
multiplying the likelihood by a Gaussian likelihood function
centered around $h=0.72$ and with a standard deviation $\sigma =
0.08$. We impose a weak Gaussian prior on the baryon density
$\Omega_b h^2 = 0.022 \pm 0.002$ (1$\sigma$) from the Big Bang
nucleosynthesis\cite{bbn}.

For CMB data, we use the three-year WMAP (WMAP-3)
Temperature-Temperature (TT) and Temperature-Polarization (TE)
power spectrum with the routine for computing the likelihood
supplied by the WMAP team\footnote{Available at
http://lambda.gsfc.nasa.gov/product/map/current/}
\cite{Spergel:2006hy,Page:2006hz,Hinshaw:2006,Jarosik:2006} as
well as ACBAR~\cite{Kuo:2002ua}, CBI~\cite{Pearson02,CBIdata} and
VSA~\cite{VSA3} data. To break the degeneracies of the
cosmological parameters, we add non-CMB data into our analysis.
For supernova type Ia (SN Ia) of ``Riess gold
sample"~\cite{Riess:2004nr}, we have marginalized over the
nuisance parameter\cite{DiPietro:2002cz} in the calculation of SN
Ia likelihood. For LSS information, we have used the 3D matter
power spectrum of SDSS\cite{Tegmark:2003uf} and 2dFGRS
\cite{Cole:2005sx}, Lyman-$\alpha$ forest data~(Ly$\alpha$) from
SDSS~\cite{lya} and the recent measurement of the baryon acoustic
oscillation (BAO) feature in the 2-point correlation function of
SDSS~\cite{Eisenstein2005}.
 To be conservative but more robust, we only use
the first 14 bins of the SDSS 3D matter power spectrum, which are
well within the linear regime\cite{sdssfit}. For Ly$\alpha$
likelihood, we modify the interpolating code\footnote{Available at
http://www.cita.utoronto.ca/~pmcdonal/LyaF/public.lyafchisq.tar.gz}
to incorporate our models. For BAO likelihood, we use the
constraint~\cite{Eisenstein2005}: \be
 A \equiv D_V(0.35)
{\sqrt{\Omega_m H_0^2}\over 0.35 c} = 0.469 \pm 0.017~~,\ee \be
D_V(z) = \left[ D_M(z)^2 {cz\over H(z)}\right]^{1/3}~~, \ee where
$H(z)$ is the Hubble parameter, $c$ is the speed of light and
$D_M(z)$ is the comoving angular diameter distance at a specific
redshift $z$.
 Moreover, the Heidelberg-Moscow experiment uses the half time of
$0\nu2\beta$ decay to constrain the effective Majorana mass and
this translates to the constraint on the sum of neutrino masses
under some assumptions\cite{DeLaMacorra:2006tu}: \be\label{HMP}
m_{\nu 0}\sim 1.8\pm0.6~eV~(2\sigma)~.\ee Given that the
Heidelberg-Moscow experiment is controversial for the time being,
we just make a tentative fit choosing the HM prior.

 For each regular calculation, we run 6
independent chains comprising of 150,000-300,000 chain elements
and spend thousands of CPU hours to calculate on a cluster. The
average acceptance rate is about 40\%. We eliminate the first
10$\%$ chain elements for ``build in", and for the convergence
test typically we get the chains satisfy the Gelman and
Rubin\cite{GR92} criteria where $R-1<0.1$.

\begin{table}
  \centering
  \caption{Mean and 1$\sigma$ constraints on the cosmological
  parameters. For the weakly constrained parameters, such as
  $m_{\nu 0}$ and $\delta$ for some data combinations, we quote the
  $95\%$ upper limits instead. Upper part of the table is for $\Lambda CDM$
  + neutrinos with constant mass while in the lower part we free $\delta$ to study
  the mass varying neutrinos. ``ALL" denotes WMAP3+ACBAR+VSA+CBI+RIESS+SDSS+2dF+Ly$\alpha$
  throughout this paper.}\label{tab}
\begin{tabular}{cccc}
  \hline
  \hline
$\delta=0$ & ALL & ALL+BAO & ALL+BAO+HM \\
$m_{\nu 0} $ (eV) & $<0.616$ & $<0.393$ & $0.760^{+0.093}_{-0.104}$ \\
$\Omega_m  $ & $0.317\pm{0.021}$ & $0.280\pm{0.015}$ & $0.303^{+0.016}_{-0.017}$ \\
$\sigma_8  $ & $0.832\pm{0.024}$ & $0.834\pm{0.024}$ & $0.795^{+0.025}_{-0.026}$ \\
   \hline
$\delta$ free & ALL & ALL+BAO & ALL+BAO+HM \\
$\delta$& $<6.661$ & $<8.562$ & $<-0.713$ \\
$m_{\nu 0}$ (eV) & $<1.619$ & $<0.776$ & $1.568^{+0.143}_{-0.141}$ \\
$\Omega_m$  & $0.319\pm{0.024}$ & $0.281\pm{0.014}$ & $0.298\pm{0.015}$ \\
$\sigma_8$   & $0.829\pm{0.027}$ & $0.835^{+0.025}_{-0.024}$ & $0.790^{+0.022}_{-0.023}$ \\
  $\Delta\chi^{2}$  & 0.198 & 0.222 & 9.752 \\
  \hline
  \hline
\end{tabular}
\end{table}

We summarize our main results of the mass varying neutrinos in the
lower part of Table I. For comparison, we also study the
$\Lambda$CDM model with neutrinos of constant mass. The first
discovery is that the neutrino mass limit at present epoch can be
relaxed dramatically if neutrino mass varies during evolution.
Without Heidelberg-Moscow data, we find that the neutrino mass
limit can be relaxed by a factor of 2.6 for ``All" data and 1.8
for ``All+BAO"\footnote{It's true that the bounds on the neutrino
mass at current epoch can be relaxed by considering the
correlation among neutrino mass with dark energy
parameters\cite{DE-nu} and with inflationary
parameters\cite{DE-run}, however if the neutrino mass varies with
time, its mass limit can be much looser. }. Adding
Heidelberg-Moscow prior, we find that the mean value of neutrino
mass rises up from 0.760 to 1.568 while reducing $\chi^{2}$ by
9.752. This is expected since the Heidelberg-Moscow prior is in
great tension with the cosmological observations. For example, the
authors of Ref.\cite{Seljak:2006bg} argued that given the current
cosmological constraint on the (constant) neutrino mass the HM
prior can be excluded. However, this controversy can be resolved
if neutrino mass varies. For our \emph{ad hoc} parametrization
(\ref{para}), we see that the ``All+BAO+HM" prior can put
stringent limit on $\delta$, namely, $-1<\delta<-0.713$. The best
fit value of $\delta$ is about $-0.9$. It means that, in order to
be consistent with all the cosmological observations, the neutrino
mass must be very small in the past, but has grown recently in
order to agree with the Heidelberg-Moscow data.

\begin{figure}[htbp]
\begin{center}
\includegraphics[scale=0.4]{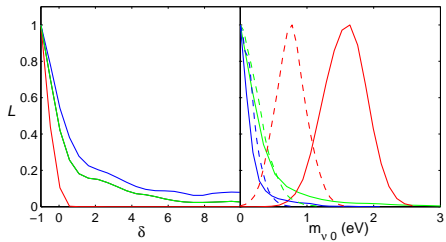}
\caption{One dimensional posterior distribution of neutrino mass
today $m_0$ and its time variation $\delta$. Solid curves denotes
the mass varying models (see text) while dashed lines show the
models with constant neutrino mass. Different data combinations
are distinguished by color. Green: All; Blue: All+BAO; Red:
All+BAO+HM prior. \label{fig:fig1}}
\end{center}
\end{figure}

\begin{figure}[htbp]
\begin{center}
\includegraphics[scale=0.4]{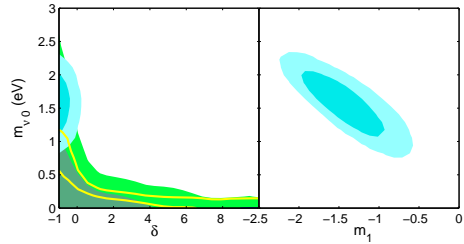}
\caption{Contour plots of parameters related to mass varying
effect of neutrino. $m_1$ is the neutrino mass changing with time,
say, $m_1=m_0*\delta$. Different data combinations are
distinguished by color. Green: All; Yellow: All+BAO; Cyan:
All+BAO+HM prior. $68\%$ and $95\%$ C.L. contours are illustrated
from inside out.\label{fig:fig2}}
\end{center}
\end{figure}

Mass varying neutrinos lessen the tension between the HM prior and
the cosmological data thus are more favored at nearly 3$\sigma$
than neutrinos with constant mass. These results are shown
graphically in Fig.\ref{fig:fig1}. From the left panel we see that
$\delta$ is weakly constrained unless we add HM prior while the
right panel shows explicitly the modification of posterior
distribution of neutrino mass today if we allow it to vary with
cosmic time. Furthermore, from Fig.\ref{fig:fig2} we see that the
current neutrino mass is correlated with its time variation. This
correlation mainly stems from the LSS data. We know that the
galaxy survey is powerful to weigh neutrinos by detecting the
suppression on small scales due to the free streaming effect of
neutrinos\cite{Hu:1997mj}. The free streaming scale of mass
varying neutrinos $\lambda_{FS-\nu}$ can be roughly estimated as:
\begin{equation}\label{fs}
    \lambda_{FS-\nu}\simeq 20  (\frac{m_\nu (a_{NR})}{30
    eV})^{-1}~Mpc~~,
\end{equation}
where $a_{NR}$ is the scale factor when neutrinos become
non-relativistic. We have seen from Eqs.(\ref{fs}) and
(\ref{para}) that $\lambda_{FS-\nu}$ is determined by the neutrino
mass today and its evolution behavior, which lead to the
correlation among $m_{\nu 0}$, $\delta$ and $m_{\nu 0}*\delta$. In
Fig.\ref{fig:fig2} we find that $m_{\nu 0}$ is anti-correlated
with $\delta$ and $m_{\nu 0}*\delta$ as expected.

\begin{figure}[htbp]
\begin{center}
\includegraphics[scale=0.4]{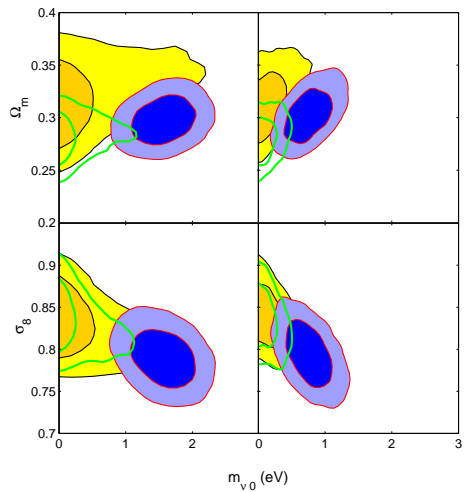}
\caption{Contour plots of sum of neutrino mass at present versus
$\Omega_m$ and $\sigma_8$. Left panel: Neutrino mass varies as
Eq.{\ref{para}}; Right panel: Constant neutrino mass. Different data
combinations are distinguished by color. Yellow: All; Green:
All+BAO; Blue: All+BAO+HM prior. $68\%$ and $95\%$ C.L. contours are
illustrated from inside out.\label{fig:fig3}}
\end{center}
\end{figure}

From Table \ref{tab}, we find that the mean value and the error
bars of $\Omega_m$ and $\sigma_8$ do not change much if neutrino
mass varies. At first glance this seems at odds since the neutrino
mass limit today has been significantly relaxed by its time
variation and we know that the neutrino mass today is strongly
correlated with matter density and $\sigma_8$ as illustrated in
Fig.(\ref{fig:fig3}). However since $\delta$ is anti-correlated
with $m_{\nu 0}$, the aforementioned effect is counteracted,
leaves $\Omega_m$ and $\sigma_8$ nearly unchanged. This means mass
varying neutrinos can hardly be excluded by data sensitive to
$\Omega_m$ and $\sigma_8$, such as SN Ia, CMB and LSS.

\begin{figure}[htbp]
\begin{center}
\includegraphics[scale=0.4]{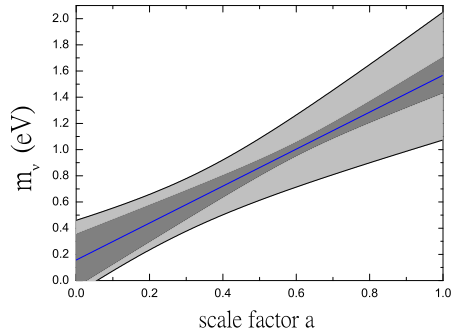}
\caption{Evolution of neutrino mass with respect to the scale
factor. From inside out, we show the models with mean parameter
value (central blue line), $68\%$ C.L.(dark gray) and $95\%$ C.L.
(bright gray) fitting with the data combination of All+BAO+HM
prior. \label{fig:fig4}}
\end{center}
\end{figure}

In Fig.(\ref{fig:fig4}), we show the evolution of neutrino mass
with 1 and 2 $\sigma$ error using all data mentioned in this
paper. we see that the neutrino mass is best measured at an
intermediate redshift rather than now due to the sensitive SN data
in this redshift range.

In summary, in this paper, we for the first time study the
cosmological implications of the time variation of neutrino mass
in a model-independent fashion. We find that numerically time
variation of neutrino mass can relax the current mass limit
significantly. This result has some interesting and important
implications. It could resolve the tension between HM prior, if
confirmed, and cosmological data, while it does not spoil the
common prediction of $\Omega_m$ and $\sigma_8$, and it might also
be possible to revive the models of warm dark matter which has
been shown to be excluded\cite{Seljak:2006qw}.

{\bf{Acknowledgments:}} We acknowledge the use of the Legacy Archive
for Microwave Background Data Analysis (LAMBDA). Support for LAMBDA
is provided by the NASA Office of Space Science. We have performed
our numerical analysis on the Shanghai Supercomputer Center(SSC). We
thank Mingzhe Li, Pei-Hong Gu, Hong Li and Xiao-Jun Bi for helpful
discussions. This work is supported in part by National Natural
Science Foundation of China under Grant Nos. 90303004, 10533010 and
19925523.

{}

\end{document}